\documentclass[sigconf]{acmart}

\usepackage{booktabs} 

\copyrightyear{2017} 
\acmYear{2017} 
\setcopyright{acmcopyright}
\acmConference{WPES'17}{October 30, 2017}{Dallas, TX, USA}\acmPrice{15.00}\acmDOI{10.1145/3139550.3139555}
\acmISBN{978-1-4503-5175-1/17/10}

\usepackage{balance}
\usepackage[normalem]{ulem}

\newcommand{\Qavg}[0]{\overline{\text{r}}} 
\newcommand{\Racc}[0]{r_{95}} 
\newcommand{\dQ}[1]{d(#1)} 
\newcommand{\dM}[1]{d_{\mathcal{P}}(#1)} 
\newcommand{\Xalph}[0]{\mathcal{X}} 
\newcommand{\Zalph}[0]{\mathcal{Z}} 

\newcommand{\Perr}[0]{p_e}
\newcommand{\Perrmin}[0]{p_e^{*}}
\newcommand{\Perravg}[0]{\overline{p}_e}

\newcommand{\geoindnospace}[0]{\textsf{GeoInd}}
\newcommand{\geoind}[0]{\textsf{GeoInd~}}
\newcommand{\Geoind}[0]{\textsf{GeoInd~}}

\graphicspath{{./img/}}

\begin{document}
\title{Is Geo-Indistinguishability What You Are Looking for?}

\author{Simon Oya}
\affiliation{%
  \institution{University of Vigo}
  \streetaddress{}
}
\email{simonoya@gts.uvigo.es}

\author{Carmela Troncoso}
\affiliation{%
  \institution{IMDEA Software Institute}
  \streetaddress{}
}
\email{carmela.troncoso@imdea.org}

\author{Fernando P{\'e}rez-Gonz{\'a}lez}
\affiliation{%
  \institution{University of Vigo}
  \streetaddress{}
}
\email{fperez@gts.uvigo.es}

\begin{abstract}
Since its proposal in 2013, geo-indistinguishability has been consolidated as a formal notion of location privacy, generating a rich body of literature building on this idea. A problem with most of these follow-up works is that they blindly rely on geo-indistinguishability to provide location privacy, ignoring the numerical interpretation of this privacy guarantee. 
In this paper, we provide an alternative formulation of geo-indistinguishability as an adversary error, and use it to show that the privacy vs.~utility trade-off that can be obtained is not as appealing as implied by the literature. We also show that although geo-indistinguishability guarantees a lower bound on the adversary's error, this comes at the cost of achieving poorer performance than other noise generation mechanisms in terms of average error, and enabling the possibility of exposing obfuscated locations that are useless from the quality of service point of view. 
\end{abstract}

\maketitle

\section{Introduction}
Geo-indistinguishability (\geoindnospace), a formal notion of location privacy introduced in~\cite{GeoInd2013CCS}, builds on the concept of differential privacy~\cite{dwork2008differential} to design user-centric location privacy-preserving mechanisms. To gain privacy while preserving some utility, in these mechanisms users report to service providers obfuscated versions of their actual locations. \geoind guarantees that obfuscated locations are statistically indistinguishable from other locations within a radius around the users' real location.
One of the most appealing features of \geoindnospace, inherited from differential privacy, is that it guarantees that, regardless of any side-information about the user she might have, the adversary learns little additional information about the real location from observing the obfuscated version.

Since its proposal~\cite{GeoInd2013CCS}, \geoind has drawn a lot of attention from the research community. A first research line extends this notion to consider mobility traces instead of single locations~\cite{Chatzikokolakis2014predictive,xiao2015protecting}, or to consider semantic and geographic privacy~\cite{Elastic2015PETS}. Some works focus on how to use \geoindnospace, or on integrating \geoind with other privacy metrics~\cite{shokri2015privacy,yu2017dynamic} to design optimal location privacy-preserving mechanisms, either in simplified~\cite{OptGeoInd2014CCS} or realistic~\cite{Practical2016} scenarios.
Finally, \geoind has been also used to implement plugins to sanitize locations for its use by other mobile applications~\cite{FawazS14} or browsers~\cite{LocationGuard}. 

A common issue in these works is that they choose \geoind based on its core \emph{qualitative} advantage, namely that it provides protection for the users in a region around their real location regardless of the adversary's side-information. However, they do not evaluate and reason \emph{quantitatively} about how much protection the mechanisms provide, i.e., if the level of privacy they achieve is meaningful.

In this work, we illustrate that \geoind can be misleading both in terms of privacy and utility. We propose an alternative definition of this privacy notion as an adversary's error, and study numerically the privacy level provided by the state-of-the-art mechanisms that guarantee this property. We also examine the trade-off between privacy and utility, showing that even though \geoind mechanisms ensure a minimum privacy protection, this comes at the expense of performing poorly in terms of average protection, and possibly generating an obfuscated location very far away from the user. 

\section{Geo-Indistinguishability}
\label{sec:preliminaries}

We first describe the operation of user-centric perturbation-based sporadic location privacy mechanisms. Consider a user, Alice, that wants to get some service from a service provider from her real location $x\in\Xalph$. Before exposing her location to the provider, Alice uses a location privacy mechanism $f$ to generate an obfuscated location $z\in\Zalph$, with probability $f(z|x)$. $\Xalph$ and $\Zalph$ are sets of locations that we assume discrete for notational simplicity, although we note that all the results in this paper are applicable to the continuous scenario.
By using mechanism $f$, Alice trades in utility for privacy. For example, if Alice's query is ``give me the bars in a radius of $100$ meters from my location'', releasing an obfuscated location $z$ away from $x$ might result in bars that are far away from her, but also protects her location since the probabilistic nature of the mechanism $f$ prevents the adversary from learning her true location $x$.

We define the \emph{multiplicative distance} between two distributions $\sigma_1(s)$ and $\sigma_2(s)$ on a set $\mathcal{S}$ as
$\dM{\sigma_1(s),\sigma_2(s)}\doteq\text{sup}_{s\in\mathcal{S}} \left|\log\frac{\sigma_1(s)}{\sigma_2(s)}\right|$
with the convention that $\left|\log\frac{\sigma_1(s)}{\sigma_2(s)}\right|$ is 0 if $\sigma_1(s)=\sigma_2(s)=0$ and $\infty$ if only one of the two is 0.

In this scenario, geo-indistinguishability is defined as~\cite{GeoInd2013CCS}:
\begin{definition}[$\epsilon$-Geo-Indistinguishability]
  A mechanism $f$ provides $\epsilon$-geo-indistinguishability if and only if, for all input locations $x,x'\in\Xalph$, the following holds
 \begin{equation} \label{eq:geoind}
   \dM{f(z|x),f(z|x')} \leq  \epsilon\cdot\dQ{x,x'}\,,
 \end{equation}
 where $\dQ{x,x'}$ is the Euclidean distance between $x$ and $x'$.
\end{definition}

The rationale behind this privacy notion is the following: by bounding the multiplicative distance, we ensure that the probability that Alice reports $z$ when she is in $x$ is similar to the probability that she reports $z$ when she is in $x'$ (up to a multiplicative factor of $e^{\epsilon\cdot\dQ{x,x'}}$). Therefore, an adversary observing $z$ cannot statistically distinguish between $x$ and $x'$ as Alice's real location. The upper bound in \eqref{eq:geoind} depends on $\dQ{x,x'}$ and $\epsilon$. The former dependence is very intuitive: given an obfuscated location $z$, two locations $x,x'\in\Xalph$ that are very close result harder to distinguish (i.e., $f(z|x)$ is close to $f(z|x')$) than if they were further apart. The role of $\epsilon$, on the other hand, is to tune the degree of \geoindnospace. Smaller values of this parameter ensure that $f(z|x)$ and $f(z|x')$ are closer, and therefore provide a higher degree of privacy than larger values.

\vspace{0.1cm}\noindent\ \textbf{Prior-Agnostic Protection.}
\geoind provides a privacy guarantee independent of any side information about $x$ the adversary might have. Let $\pi(x)$ be a probability mass function over $x\in\Xalph$ representing the \emph{prior} adversary's side information about Alice's real location $x$. After observing $z$, the adversary can update her knowledge by computing the posterior probability mass function
\begin{equation} \label{eq:posterior}
 p(x|z)=\frac{ f(z|x)\cdot \pi(x)}{\sum_{x'\in\Xalph} f(z|x') \cdot \pi(x')}\,.
\end{equation}

By using \eqref{eq:geoind} and \eqref{eq:posterior}, it is easy to show that \geoind implies
\begin{equation}
 \dM{p(x|z),\pi(x)} \leq \epsilon\cdot d(\pi)\,,
\end{equation}
where $d(\pi)$ is the maximum distance between two locations $x$ and $x'$ such that $\pi(x)>0$ and $\pi(x')>0$. In other words, \geoind ensures a certain degree of similarity between the adversary's prior and posterior information about Alice's real location, for any prior $\pi$. Note that \geoind is not an absolute privacy guarantee, but only ensures that given $z$ the adversary gets no significant \emph{extra} accuracy with respect to the prior. However, if given this prior the adversary can pinpoint a user's location to a small region in the map (small $d(\pi)$), then even though $z$ does reveal little information about $x$, the adversary's estimation of $x$ will still be accurate. 

\vspace{0.1cm}\noindent\ \textbf{Choosing the Privacy Parameter.}
The general approach to selecting a proper value for the parameter $\epsilon$ is to pick a \emph{privacy level} $\epsilon^*$ and a \emph{privacy radius} $r^*$, and set $\epsilon=\epsilon^*/r^*$. This ensures that, when Alice is in $x$ and releases $z$, her location is statistically indistinguishable from all the other locations $x'$ within a radius of $r^*$ around her, i.e., $\dM{f(z|x),f(z|x')}\leq \epsilon^*$ as in \eqref{eq:geoind}.

Quantitatively, however, it is hard to determine if a bound on the multiplicative distance $\epsilon^*$ gives ``enough privacy''. This is reflected in the literature, where there is no consensus about which value of the bound in \eqref{eq:geoind} denotes a high degree of indistinguishability. In the seminal paper~\cite{GeoInd2013CCS}, Andr{\'e}s et.~al choose $\epsilon^*=\log2$ in a radius of $r=200$ meters as the highest privacy level. This bound is used by some follow-up works~\cite{LocationGuard,Elastic2015PETS}, while others take different values of $\epsilon^*$, ranging from $\epsilon^*=\log10$~\cite{Chatzikokolakis2014predictive} to $\epsilon^*=\log1.4$~\cite{Practical2016}.

\section{\Geoind as an Adversary Error}
\label{sec:perr}
In this section, we introduce an alternative characterization of \geoind as an adversary error. This characterization helps us in providing more intuition behind the privacy level obtained for a specific value of the privacy parameter $\epsilon$, and in understanding the protection it provides beyond the upper bound expressed in \eqref{eq:geoind}.

Consider that the adversary's side information is that Alice is equally likely in either of two locations $x$ and $x'$, i.e., $\pi(x)=\pi(x')=0.5$. After observing $z$, the adversary has to decide between $x$ and $x'$. We refer to this adversary as the \emph{decision adversary}. Assume, without loss of generality, that $f(z|x)\geq f(z|x')$, and thus the optimal decision in terms of minimizing the adversary's probability of error is deciding that Alice's location is $x$. In this case, the adversary's probability of error is
\begin{equation} \label{eq:Perr}
 \Perr(x,x',z)=\frac{f(z|x')}{f(z|x)+f(z|x')}\,.
\end{equation}
Then, Geo-indistinguishability can be defined as follows:
\begin{lemma}[$\epsilon$-Geo-Indistinguishability as error]
 A mechanism $f$ guarantees $\epsilon$- geo-indistinguishability \emph{if and only if}, for any pair of input locations $x,x'\in\Xalph$ and any output location $z\in\Zalph$, it ensures that the minimum probability of error $\Perrmin$ of the decision adversary described above is
 \begin{equation} \label{eq:Perrmin}
 \Perr(x,x',z)\geq \Perrmin=\frac{1}{1+e^{\epsilon\cdot\dQ{x,x'}}}\,.
\end{equation}
\end{lemma}
It is easy to see that the \geoind definitions in \eqref{eq:geoind} and \eqref{eq:Perrmin} are equivalent by substituting \eqref{eq:Perr} in \eqref{eq:Perrmin} and operating. We have chosen $\pi(x)=\pi(x')=0.5$ as prior knowledge for the decision adversary, as this ensures the guarantee in \eqref{eq:geoind}, but we note that when $\pi(x)\neq\pi(x')$ \geoind does not guarantee a minimum probability of error against this adversary.

This alternative definition of \geoind allows us to intuitively interpret the privacy guarantee achieved by $f$. For example, given $\epsilon=2\text{km}^{-1}$ and two locations $x$ and $x'$ separated $\dQ{x,x'}=0.5$km, according to \eqref{eq:geoind} the multiplicative distance between $f(z|x)$ and $f(z|x')$ is bounded by $1$. However, whether an upper bound on the multiplicative distance of $1$ is a reasonable level of protection is not clear. In terms of probability of error, \eqref{eq:Perrmin} bounds the adversary's error to be $\Perr\geq\Perrmin=0.27$: before observing $z$ the decision adversary has a probability of correctly guessing Alice's input location of $0.5$, and after the release her probability of success is in the worst case $0.73$. It is difficult to consider this worst case probability as ``high indistinguishability'', which contradicts the \geoind idea of achieving indistinguishability for every pair of input locations. We explore the implications of this interpretation in the next section.

\section{\Geoind in Numbers}
\label{sec:evalnum}
In this section, we quantitatively evaluate the privacy and utility achieved by \geoind mechanisms. For privacy, given a fixed distance between two locations $x$ and $x'$, we measure both the upper bound on the multiplicative distance in \eqref{eq:geoind}, $\epsilon^*=\epsilon\cdot\dQ{x,x'}$, and the lower bound on the probability of error \eqref{eq:Perr} of the decision adversary, $\Perrmin$. For utility we consider two metrics: the average loss $\Qavg$, measured as the average Euclidean distance between the actual location of the user $x$ and the obfuscated location $z$~\cite{OptGeoInd2014CCS,xiao2015protecting,shokri2015privacy,Elastic2015PETS,Practical2016,yu2017dynamic}, and the radius of the circular region centered around $x$ where $z$ is with probability $0.95$, denoted by $\Racc$~\cite{LocationGuard}.

We evaluate two \geoind mechanisms. First, the planar Laplace mechanism, proposed in the seminal work~\cite{GeoInd2013CCS}, implemented in~\cite{FawazS14,LocationGuard}, and used as a baseline for comparison in~\cite{OptGeoInd2014CCS,Practical2016}. This mechanism generates obfuscated locations $z$ from the actual location $x$ by adding 2-dimensional Laplace noise to the latter. Second, the planar Laplace with remapping~\cite{Practical2016}, current state-of-the-art. This mechanism first generates a temporary location $z'$ by adding 2-dimensional Laplace noise to $x$, and then performs a deterministic remapping from $z'$ to $z$ that is designed to minimize the average loss $\Qavg$ of the scheme while providing the same privacy guarantees as the original version. This remapping is computed using a dataset with information about the popularity of each input location $x\in\Xalph$, and therefore the mechanism is tied to a particular dataset. We leave optimal mechanisms that can only be implemented in simple discrete scenarios \cite{OptGeoInd2014CCS,shokri2015privacy,yu2017dynamic} out of our evaluation, as the low-resolution quantization needed to implement them in a real scenario makes them suboptimal (cf.~\cite{Practical2016}).

\begin{figure}[t]
  \centering
  \includegraphics[width=0.8\columnwidth]{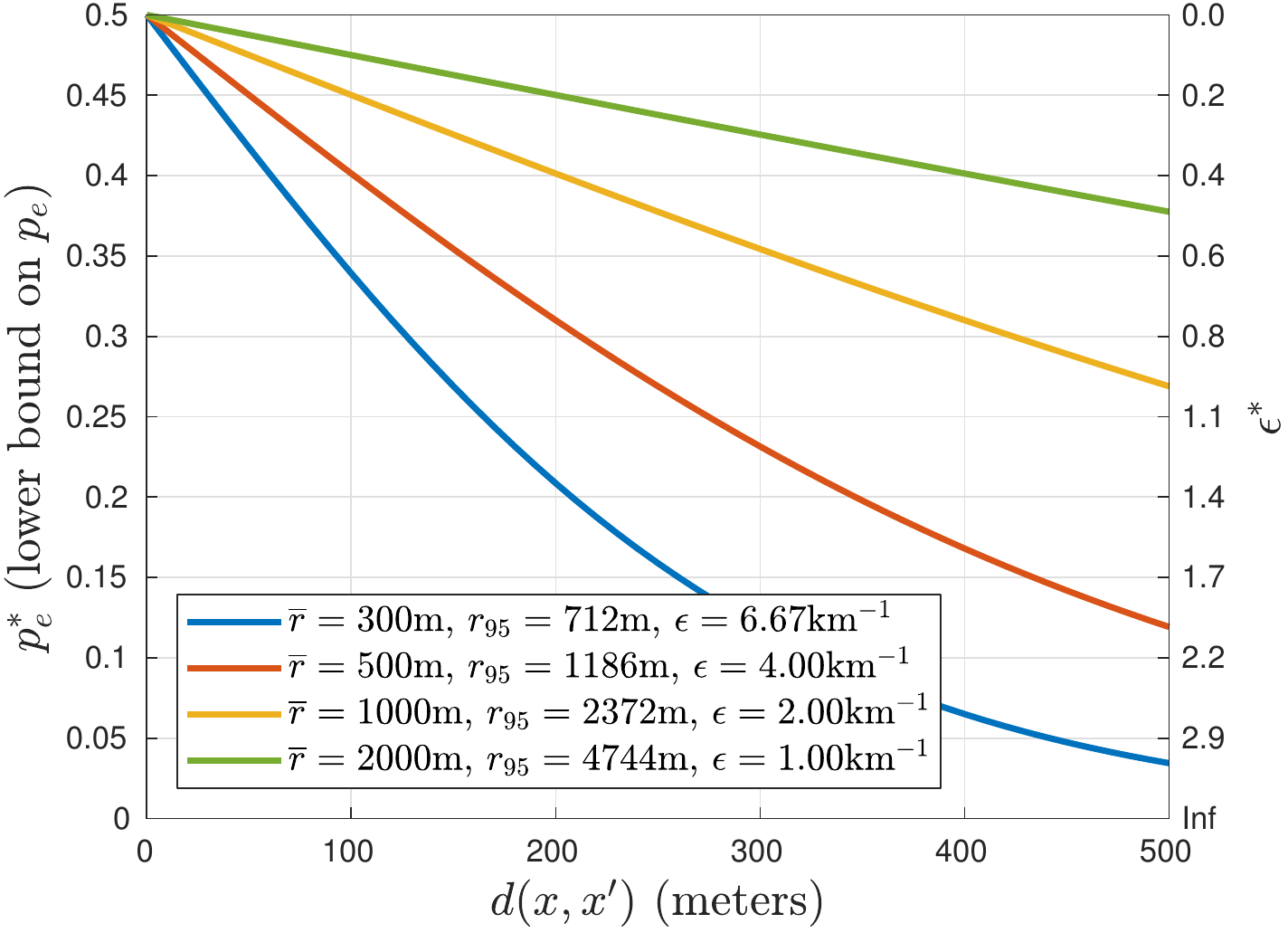}
  \caption{Performance of the Planar Laplace mechanism.}
  \label{fig:res_Laplace}
\end{figure}

\vspace{0.1cm}\noindent\ \textbf{Planar Laplace Mechanism.}
Given a privacy value $\epsilon$, the average loss of the Laplace mechanism is $\Qavg=2/\epsilon$, and $\Racc$ can be computed analytically using the Lambert W function (cf.~\cite{GeoInd2013CCS}). The value of $\Perrmin$ can be computed from $\epsilon$ following \eqref{eq:Perrmin}. We show $\Perrmin$ and $\epsilon^*$ for this mechanism, when locations are separated $\dQ{x,x'}$ meters, in Figure~\ref{fig:res_Laplace} for different utility levels. As expected, as we add more noise (larger $\Qavg$ or $\Racc$), protection improves (larger $\Perrmin$ or smaller $\epsilon^*$). 

To better understand the trade-off between privacy and utility let us consider as reference a privacy level $\Perrmin=0.4$ (i.e., the decision adversary succeeds at most 60\% of the times). To obtain such protection level in a radius of $r^*$ one needs to add Laplacian noise with average loss of $\Qavg\approx 5 r^*$. This results $5\%$ of the time on an obfuscated location $z$ further than $r_{95}\approx 12 r^*$ from the real location $x$. Consider that we want $\Perrmin=0.4$ in locations within a radius of $r^*=200$m. In this case, the obfuscated location would be on average $\Qavg=1$km away from the real location, and $5\%$ of the time it would be further than $2.3$km away (yellow line in Fig.~\ref{fig:res_Laplace}). In applications that are not sensitive to large amounts of noise (e.g., weather forecast) this might be reasonable. However, in other applications where one would require a utility in the same order of magnitude as the privacy protection (e.g., finding nearby points of interest), the Laplace mechanism and, up to some extent, \geoindnospace, are not desirable.

\vspace{0.1cm}\noindent\ \textbf{Planar Laplace with Optimal Remapping.}
Since this mechanism cannot be evaluated analytically, we follow the empirical approach in \cite{Practical2016}: we use $80\%$ of the users from Gowalla dataset\footnote{\url{https://snap.stanford.edu/data/loc-gowalla.html}} to design the remapping function, and use the remaining $12\,112$ users as a testing set to evaluate the utility of the mechanism after remapping. We generate an output $z$ for $20\,000$ random checkins from testing set users, for values of $\epsilon$ from the previous experiment ($\epsilon=\{6.67, 4, 2, 1\}$, in km$^{-1}$), and use them to compute $\Qavg$ and $\Racc$. The results in terms of $\Perrmin$ and $\epsilon^*$ vs. $\dQ{x,x'}$ coincide with the ones in Fig.~\ref{fig:res_Laplace}, but we obtain much better quality: $\Qavg=159$, $266$, $578$ and $1271$ meters, i.e., $37-47\%$ smaller than plain Laplace. The $95\%$ loss percentile in each case is $\Racc=565$, $999$, $2146$, and $4162$ meters, which is only a $10-21\%$ reduction from the planar Laplace without remapping. To obtain a protection of $\Perrmin=0.4$ in a radius of $r^*$ around the real location, in this scenario one needs to add noise with a loss of roughly $\Qavg\approx 3 r^*$ and $\Racc\approx 10 r^*$. Although the average loss reduction is considerable, the utility cost is still large compared to the radius of the privacy region this mechanism ensures. This highlights the importance of analyzing 
\geoind numerically to understand the actual privacy vs.~utility trade-off it provides.

\begin{figure*}[ht]
  \begin{minipage}[b]{0.41\linewidth}
    \centering
    \includegraphics[width=\columnwidth]{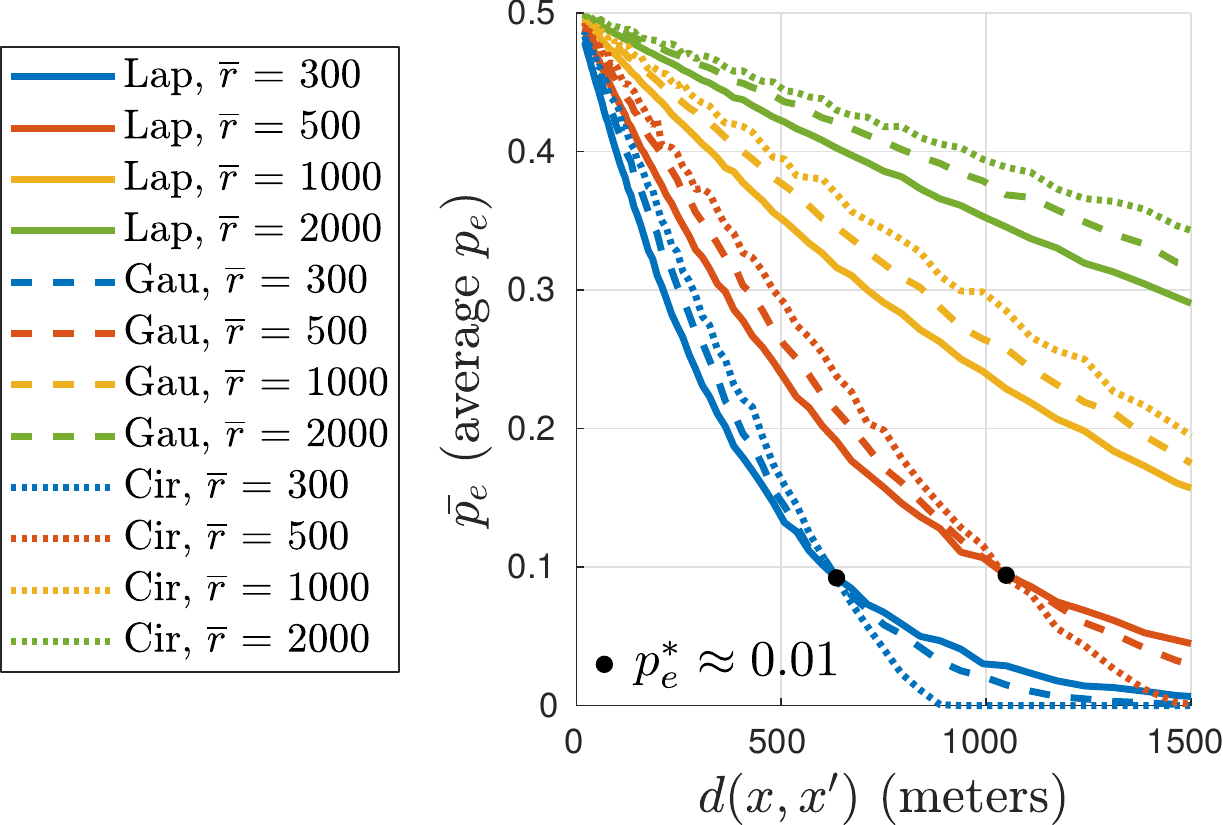}
    \caption{Performance in terms of the average probability of error of the decision adversary.}
    \label{fig:avg_pe_mechanisms}
  \end{minipage} \hfill
  \begin{minipage}[b]{0.3\linewidth}
    \centering
  \includegraphics[width=\columnwidth]{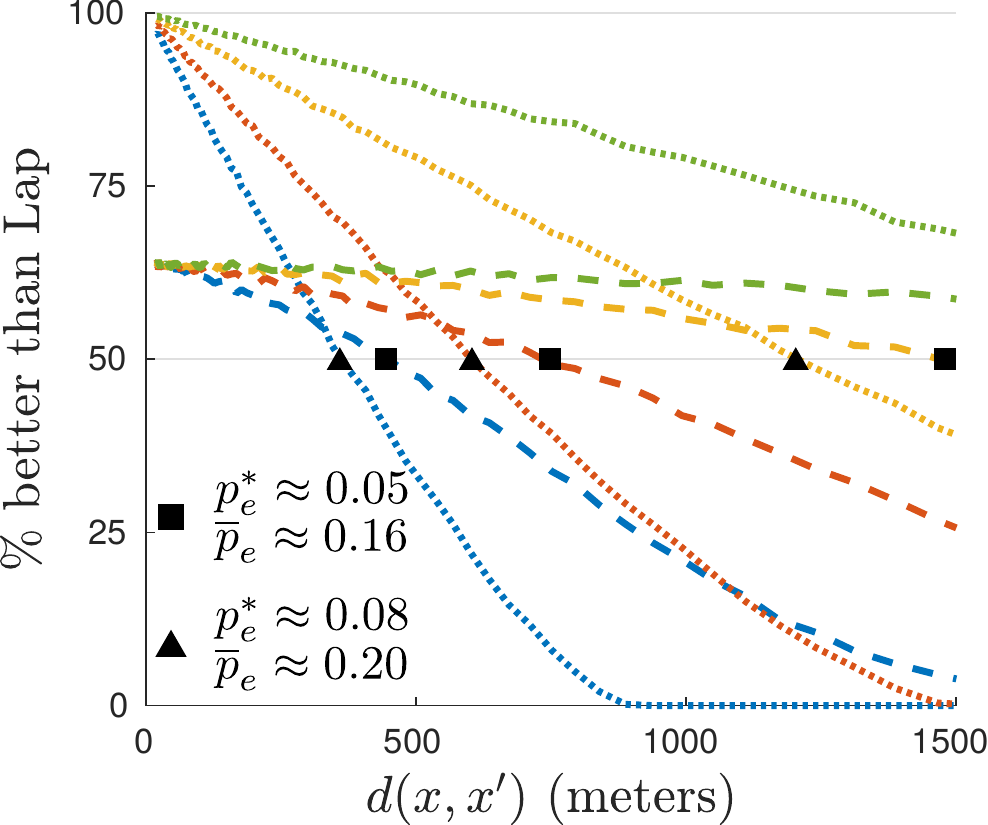}
  \caption{Percentage of times the Gaussian and Circular mechanisms outperform the Laplace mechanism.}
  \label{fig:PCT_better}
  \end{minipage} \hfill
  \begin{minipage}[b]{0.24\linewidth}
    \centering
  \includegraphics[width=\columnwidth]{{{NormHist_Q0.5}}}
  \caption{Normalized histogram of $\Perr$, for $\dQ{x,x'}\approx 100$ and $\Qavg=500$ meters ($\Perrmin=0.4$).}
  \label{fig:histogram}
  \end{minipage} \hfill
\end{figure*}

\section{Other Properties of \Geoind}
\label{sec:evalconcept}

So far we have studied the lower bound ($\Perrmin$) \geoind mechanisms provide on the probability of error of the decision adversary ($\Perr$). We now study other properties of \geoind mechanisms against this adversary. For this purpose, we evaluate three mechanisms: the planar Laplace mechanism, described above, and the Gaussian and uniform circular mechanisms. The latter mechanisms generate $z$ by adding to the real location $x$, respectively, 2-dimensional Gaussian noise and uniform noise in a circle. We choose not to use remapping, as its improvement would be similar for all mechanisms and thus does not influence the comparison. For each experiment, we consider two locations $x$ and $x'$ separated a distance $\dQ{x,x'}$ and generate $z$ using the three mechanisms. Then, we measure the probability of error of the decision adversary $\Perr$ in that realization, and repeat this $20\,000$ times for different values of $\dQ{x,x'}$ and $\Qavg$. 

The average probability of error of these mechanisms, denoted by $\Perravg$ and computed by averaging the $20\,000$ samples of $\Perr$, is shown in Figure~\ref{fig:avg_pe_mechanisms}. Given an average loss $\Qavg$, both Gaussian and circular mechanisms achieve a larger average error than the Laplace mechanism, up to a certain distance $\dQ{x,x'}$ marked with $\bullet$ in the figure. We do not show these marks for $\Qavg=1000$ and $\Qavg=2000$, but they also lay close to $\Perravg=0.1$. At these points, the Laplace mechanism achieves $\Perrmin\approx 0.01$ and $\Perravg\approx 0.1$ for all tested values of $\Qavg$. The fact that the Laplace mechanism performs better from $\bullet$ onwards is not significant: in these scenarios, regardless of the mechanism, the adversary guesses the right location with an average probability larger than $0.9$, i.e., no mechanism provides privacy. We conclude that, in all relevant scenarios (i.e., reasonable privacy levels), the Gaussian and circular mechanisms achieve a larger average error than the Laplace mechanism. This means that using \geoind as a way of providing an \emph{average protection level} against an adversary with unknown side-information \cite{shokri2015privacy,yu2017dynamic} is not recommended, as it is not the guarantee that this notion provides.

Figure~\ref{fig:PCT_better} shows the percentage out of the $20\,000$ realizations where the Gaussian and circular mechanisms achieve a larger $\Perr$ than the Laplace mechanism.
We see that these mechanisms achieve a larger probability of error more often than the Laplace mechanism when $x$ and $x'$ are separated up to a distance $\dQ{x,x'}$ corresponding to the points marked with $\blacksquare$ (Gaussian) and $\blacktriangle$ (circular). The figure also shows the performance of the Laplace mechanism in terms of $\Perrmin$ and $\Perravg$ at these points (these values remain almost constant when changing $\Qavg$). Similarly to the previous case, these values represent a very low privacy regime, i.e., for all reasonable privacy levels, the Gaussian and circular mechanisms are more likely to achieve a larger $\Perr$ than the Laplace mechanism. This is better illustrated in Figure~\ref{fig:histogram}, which shows the normalized histogram of the probability of error $\Perr$ provided by the Gaussian and Laplace mechanisms for $\dQ{x,x'}=100$m and $\Qavg=500$m. As expected, the Laplace mechanism ensures a minimum probability of error ($\Perrmin=0.4$), but is not able to achieve large probabilities of error as often as the Gaussian mechanism. These experiments reinforce the idea that \geoind is not a ``cure-all'' privacy guarantee against a prior-agnostic adversary.

All the above experiments compare mechanisms offering the same average loss $\Qavg$. In terms of $\Racc$, the Laplace mechanism ($\Racc\approx2.37\Qavg$) performs worse than the circular ($\Racc\approx1.46\Qavg$) and the Gaussian ($\Racc\approx1.95\Qavg$) alternatives. Thus, compared with a fixed $\Racc$, the Laplace mechanism would perform \emph{even worse} than the others.

\section{Where to go now}
\label{sec:discussion}

Geo-indistinguishability, which provides differential privacy-like guarantees in the location privacy scenario, has drawn a lot of attention from the community. However, our quantitative evaluation shows that the (worst-case or average) privacy guarantees it provides are unsatisfying unless utility is sacrificed. The main reason for this poor performance is that, in the counting queries on a database scenario where differential privacy was initially proposed~\cite{dwork2008differential}, queries have \emph{low sensitivity}, i.e., the contribution of a single user does not significantly affect the outcome. This enables the achievement of a high privacy level (e.g., $\epsilon^*=0.01$) without introducing much noise, thus preserving utility. In the location scenario where Geo-indistinguishability operates, each query has high sensitivity and therefore requires large noise to provide protection. For instance, to achieve \geoind with $\epsilon^*=0.01$ between locations in an area of $100$m, the average loss is 20km.
Moreover, \geoind can only be achieved at the expense of having an \emph{unbounded maximum quality loss} since the guarantee \eqref{eq:geoind} no longer holds if the mechanism is truncated to ensure a minimum utility for the users.

This does not mean that \geoind should be abandoned, but we argue that it should be carefully configured, understanding the type and amount of protection it provides. Our \geoind characterization as an adversary error should assist in this task, as it helps to quantitatively interpret the degree of protection provided. We have also shown that in some scenarios there are levels of protection that are not achievable without unreasonable utility loss. Potential solutions could be to use bandwidth as a resource to improve utility~\cite{GeoInd2013CCS}, or re-design location queries to have lower sensitivity (e.g., aggregating queries \cite{dwork2008differential} locally, at the user level).

\begin{acks}
\small
This work is partially supported by EU H2020-ICT-10-2015 NEXTLEAP (GA n 688722), the Agencia Estatal de Investigaci{\'o}n (Spain) and the European Regional Development Fund (ERDF) under projects WINTER (TEC2016-76409-C2-2-R) and COMONSENS (TEC2015-69648-REDC), and by the Xunta de Galicia and the European Union (ERDF) under projects Agrupaci{\'o}n Estrat{\'e}xica Consolidada de Galicia accreditation 2016-2019 and Red Tem{\'a}tica RedTEIC 2017-2018. 
Simon Oya is funded by the Spanish Ministry of Education, Culture and Sport under the FPU grant.
\end{acks}

\bibliographystyle{ACM-Reference-Format}
\balance
\bibliography{bibliolocation} 

\end{document}